\documentclass[12pt,a4paper]{article}  % Use article class; adjust options as needed

\usepackage[utf8x]{inputenc}  % For UTF-8 encoding
\usepackage[T2A]{fontenc}     % Better font encoding
\usepackage{cite}
\usepackage{color}
\usepackage{amsmath}        % For math equations
\usepackage{graphicx}       % For including figures
\usepackage{float}          % For figure placement
\usepackage{natbib}         % For citations (use \cite{key})
\usepackage{hyperref}       % For hyperlinks in references
\usepackage{geometry}       % Page margins
\usepackage{authblk} 
\usepackage[inkscapelatex=false]{svg}

\title{Axion electron-electron interaction in the RaOH molecule to search for Dark matter}

\author[1,*]{Mikhail Reiter}
\author[1,2,$\dagger$]{Anna Zakharova}

\affil[1]{Saint-Petersburg State University,
199034, Russia, Saint-Petersburg, Universitetskaya nab. 7–9}

\affil[2]{Petersburg Nuclear Physics Institute named by B.P. Konstantinov of National Research
Centre "Kurchatov Institute", 1, mkr. Orlova roshcha, Gatchina, 188300, Russia}

\affil[*]{mikh.reiter@gmail.com}
\affil[$\dagger$]{zakharova.annet@gmail.com, anna.zakharova@spbu.ru}
\date{}

\begin{document}

\maketitle

% Abstract
\begin{abstract}
Axions are promising candidates for the role of Dark matter particles. In this paper, the question of the suitability of the RaOH molecule for the experimental detection of electron-electron interactions through the exchange of axions is studied. To take into account an impact of the rotations and vibrations a computation must be performed for a large number of molecular configurations. In this work we study this problem using a combination of the Generalized relativistic effective core potential and One-center restoration technique of the correct four-component spinors.
\end{abstract}

% Keywords (optional)

% Introduction
\section{Introduction}

The existence of Dark matter is a fact well-established by astrophysical observations, which goes beyond the physics of the Standard Model of Interactions. Its introduction makes it possible to correctly describe the velocities of stars in galaxies and explain anomalous gravitational lensing data, for example, obtained during the study of the collision of the Bullet Cluster of galaxies~\cite{Clowe_2004}. In addition, the concept of Dark matter makes it possible to describe the temperature fluctuations of cosmic microwave background radiation.~\cite{1995ApJS}. The search for the particles that make up Dark matter is one of the most important tasks facing modern theoretical and experimental physics.~\cite{chadha2022axion}. 

%Such searches were carried out (at the collider, direct searches at detectors, search for decay products of such particles) \textcolor{red}{TODO}.

As a candidate for the role of a Dark matter particle, we consider the axion, a pseudoscalar particle with a nonzero mass $m_{\rm ax}$ ~\cite{cosmo:1983, abbott1983cosmological,dine1983not}. It should be noted that in addition to solving the mystery of Dark matter, the detection of an axion in a certain mass range would  solve the strong $\mathcal{CP}$ problem of quantum chromodynamics (QCD) ~\cite{peccei1977cp}, where $\mathcal{C}$ denotes charge conjugation, $\mathcal{P}$ denotes spatial reflection.
Hypothetical pseudoscalar particles having similar interactions, but not solving the strong $\mathcal{CP}$ problem are known as axion-like particles (ALPs). To explain Dark Matter, these must be weakly interacting scalar particles in a fairly wide range of masses from $10^{-12}$ to $10^{3}$ eV. 

Later in the paper, we will use the word "axion" for both classes of particles: the ALP and QCD axions.

One of the possible methods of searching for "signals" of dark matter is to study energy shifts in the spectrum of molecules. The consequence of the existence of axions may be a number of effects that violate the conservation of $\mathcal{P}$ and $\mathcal{T}$ parities: the odd electron-nuclear and electron-electron interactions transmitted through the exchange of axions, the interaction of the electron shell with the cosmic background of axions \cite{zakharova2025hexatomic}, and contributions to the electric dipole moments of particles induced by radiational effects\cite{ourRaOH,zakharova2021rovibrational,zakharova2022impact,zakharova2022rotating,zakharova2024symmetric}. It is proposed to investigate such effects in the same experiments that were previously carried out to search for the electric dipole moment of an electron (eEDM).

The most successful experiment at the moment is the determination of eEDM in the molecule HfF$^+$~\cite{PhysRevLett.126.171301}, which provides the current limit on this effect. This is just one of the effects associated with the violation of $\mathcal{P}$, $\mathcal{T}$ -- symmetries.  In addition, it is important to note a number of works devoted to the study of other effects of parity violation: scalar-pseudoscalar electron-nuclear interaction \cite{ourRaOH, zakharova2021rovibrational,zakharova2022rotating,zakharova2024symmetric,chubukov2025t} and \textcolor{black}{anapolar moment of the nucleus ~\cite{Blanchard_2023}, the magnetic quadrupole moment of the nucleus ~\cite{PhysRevA.100.032514, 10.1063/5.0028983}
% (arXiv:1906.11487 , https://arxiv.org/abs/2006.03848) 
and other $\mathcal{CP}$-violating interactions within the nuclei~\cite{D0CP01989E, annurev:/content/journals/10.1146/annurev-nucl-121423-101030}}. 
% (https://arxiv.org/abs/2003.10885)(https://arxiv.org/abs/2501.02744) }.

%почему он аномальный дипольный момент? Лучше просто дипольный Сначлла вводим NE-SPS, а потом уже обозначение
%Здесть надо написать подробнее про P,T-нечетные эффекты, чтобы дать ссылки на статьи. 
% (\cite{ourRaOH} \cite{zakharova2021rovibrational} \cite{zakharova2022impact} da, eto tak, vishe nado dopisat', ne sharuy)
Triatomic molecules are a promising candidate for searching for Dark matter due to the presence of closely spaced energy levels of opposite parity in their spectrum, which arise from the transverse vibrations of the ligand. Such a system of closely spaced levels is called an $l$-doublet. The latter fact makes them ideal candidates for searching for effects leading to a violation of $\mathcal{P}$ and $\mathcal{T}$-parity, which includes electron-electron interaction through axion exchange. In an external electric field, the energy difference between closely spaced vibrational-rotational levels with opposite parity will be proportional to the so-called sensitivity parameter $E_{\rm ax}$

This splitting increases rapidly with increasing atomic number of the element, so it is important to consider molecules with heavy atoms. These molecules include the RaOH molecule, which has been successfully~\cite{conn2025productionspectroscopycoldradioactive}
% \textcolor{red}{[https://arxiv.org/abs/2508.08368]} 
 laser cooled, rotational and spin-rotational components were obtained. This will make it possible, in the near future, to perform precision spectroscopic studies of the electronic and rovibrational structure of the energy levels of the molecule under consideration. 
As a result of the presence of opposite parity levels and the possibility of laser cooling, the study of such molecules is simplified from both experimental and theoretical points of view.

The present study provides a theoretical estimation of the sensitivity parameter of the triatomic molecule RaOH to electron-electron interaction through the exchange of axion. For other molecules, it was made in several research studies\cite{maison2021electronic, maison2021axion, prosnyak2024axion}.  Earlier, for the same system, an estimate was made for the electron-nuclear interaction through axion exchange \cite{zakharova2025parallelized, Zakharova2026}.

Due to the presence of a heavy radium atom, calculations of the electronic structure of the molecule are computationally complex. 
% из-за необходимости учета влияния колебаний и вращений, 

To simplify this task, the methods of generalized relativistic core potential (GRECP) \cite{titov1999generalized} and the method of one-center restoration (OCR) are used\cite{titov2006d} . Previosuly this technique was successfully applied to the study of spectrum perturbation by $\mathcal{P}$, $\mathcal{T}$ violating one-electron operators. However, the electron-electron interaction is a much more complicated task requiring computation of the two-electron integrals. To our knowledge, our work is the first one to use the OCR technique in such context.

% В контексте разработки вычислительных методов в данной статье мы разрабатываем эффективную процедуру вычисления двухэлектронных интегралов аксионного обмена в рамках...
% In this paper, we apply it for the first time to calculate two-electron integrals!
% У радия еще и большая октупольная деформация ядра, что может усилить P,T нечетные эффекты. Про это можно посмотреть у Исаева Тимура, по Зайцевскому можно найти)

%ну, не так всё скучно =). Не только же применяем численные методы.  we obtain the enhancement..
%Для учета влияния колебаний и вращений требуется расчет для множества конфигураций. Иза-за тяжелого атома радия это вычислительно сложно. (возмущение, обусловленное эффектами нарушения четности, растет быстро с увеличением атомного номера элемента, то для экспериментов интересны молекулы с тяжелыми атомами-куда-нибудь вставить, где написать про выбор атома. Либо тут вписать. У радия еще и большая октупольная деформация ядра, что может усилить P,T нечетные эффекты. Про это можно посмотреть у Исаева Тимура, по Зайцевскому можно найти) Для их упрощения мы используем метод орэпо + ОЦВ.  В настоящей работе мы впервые его применяем для расчета двухэлектронных интегралов! (можно это и абстакт в краткой форме)

\section{Electron-electron interaction in molecules, accompanied by axion exchange}

\begin{figure}
\centering
\includegraphics[width=0.5\linewidth]{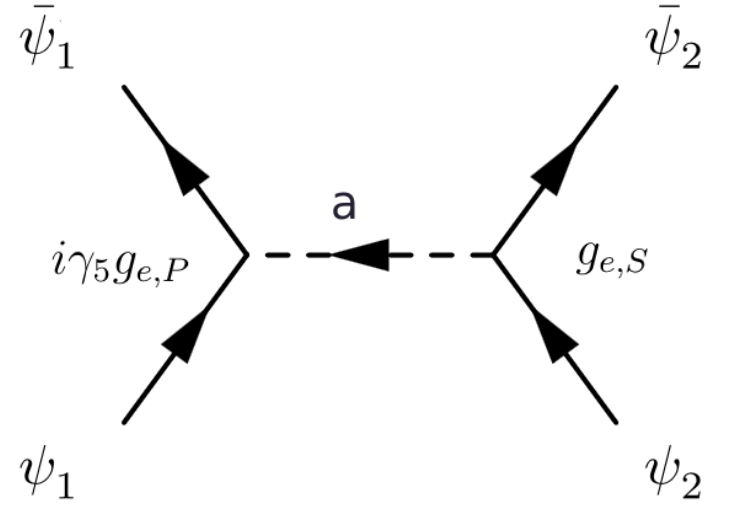}
\caption{Feynman diagram for the axion electron-electron interaction, the dotted line corresponds to the axion
\label{fig:feynman_ee}}
\end{figure}

Consider the interaction of axions (pseudoscalar field $a$) with leptons of the Standard Model $\Psi_f$. It is written with the following terms in action:

\begin{equation}
\mathcal{S}_{\rm int} = a\sum_{f\in\mathrm{SM}} \bar{\Psi}_f\Big(g_{f,S} + i g_{f,P}\gamma_5\Big) \Psi_f.
\end{equation}
Next, the electron-electron interaction will be considered, so only the coupling constants $g_{e,S}$ and $i\gamma_5 g_{e,P}$ corresponding to scalar and pseudoscalar interactions will be taken into account. As both coupling constants are expected to be extremely weak, in our work we will restrict ourselves to the tree approximation. Then the only Feynman diagram violating $\mathcal{P}$ and $\mathcal{T}$ parity is shown in Fig.~\ref{fig:feynman_ee}.

The matrix element of the $S$ matrix is calculated in the coordinate representation. A detailed output of the corresponding expressions is given in ~\cite{PRD:1984:130, PRB:2017:127}. It is important to note that the axion propagator obtained by the Fourier transformation contains a radial dependence of the Yukawa potential $\sim e^{-m_a r}/r$. We work within the approximation of $e^{-m_a r}\sim 1$, since the characteristic size of the molecule corresponds to the mass of the axion $\sim 1$ keV, which is far beyond the upper bound of the mass of the axion required to solve the strong $\mathcal{CP}$ problem~\cite{peccei1977cp}, and is also highly limited as a candidate for the role of Dark Matter particles. In the coordinate representation, the axion electron-electron interaction is expressed in terms of two-electron integrals of the form:

\begin{equation}
A(a,b|c,d) = \int d \vec{r_1} d \vec{r_2} \;\psi_a^\dagger(\vec{r_1}) (i \gamma_0 \gamma_5) \psi_c (\vec{r_1}) \; \frac{1}{r_{12}} \; \psi_b^\dagger(\vec{r_2}) \gamma_0 \psi_d (\vec{r_2}),
\label{eq:ax_two_el_int}
\end{equation}
where $r_{12} = |\vec{r_1}-\vec{r_2}|$. Note that both scalar and pseudo-scalar electron densities involve $\gamma_0$ originating from $\bar{\Psi}=\Psi^\dagger\gamma_0$.

Using these two-electron integrals, it is possible to construct the potential of electron-electron interaction:
\begin{equation}
V_{ee}(\vec{r}_1, \vec{r}_2) = \frac{g_{e,S} g_{e,P}}{4 \pi r_{12}} (\gamma_0)_1 (i\gamma_0 \gamma_5)_2, \end{equation}
where $()_i$ means the effect of the gamma matrix on the $i$-th electron. The $(\gamma_0)_1$ is often omitted in other works on this topic, however its impact becomes significant only in the vicinity of the heavy nucleus where the electrons may gain relativistic momenta.

To assess the sensitivity of the molecule in question to the detection of hypothetical axion interactions, we determine the sensitivity parameter $E_{\rm ax}$:
\begin{equation}
E_{\rm ax} = \frac{\langle \Psi_{\rm el} | \hat{H}_{ee} | \Psi_{\rm el} \rangle}{\Omega \cdot g_{e, S} \cdot g_{e, P}},
\label{eq:enh_param}
\end{equation}

In the expression above, $\Omega$ is the projection of the total electron angular momentum onto the molecular axis, and $H_{ee}$ is the Hamiltonian of $\mathcal{P}$, $\mathcal{T}$-odd interaction, defined as:
\begin{equation}
\hat{H}_{ee} = \sum_{\substack{i,j=1 \\ i \neq j}}^{N_e} V_{ee}(\vec{r}_i, \vec{r}_j),
\label{PropertyEq}
\end{equation}
where the summation is performed over all the electrons of the system. In the equation (\ref{eq:enh_param}) $\Psi_{\rm el}$ is the wave function of the electron, which depends on the geometry of the molecule. Thus, the value defined in eq.~\ref{eq:enh_param} is explicitly determined by the geometry of the molecule and must be averaged with the nuclear wave function. 

The enhancement parameter is the cornerstone of this study, since, for example, the energy shift of the levels of a $K$ doublet in an external electric field will be equal to:
\begin{equation}
E_{+M} - E_{-M} = P \cdot g_{e,S} \cdot g_{e,p} \cdot E_{\rm ax},
\end{equation}

where $P$ is the polarizability coefficient (which can be set equal to $1$ in a strong electric field; see~\cite{petrov2022sensitivity}.)

\begin{figure}
\centering
\includegraphics[width=0.5\linewidth]{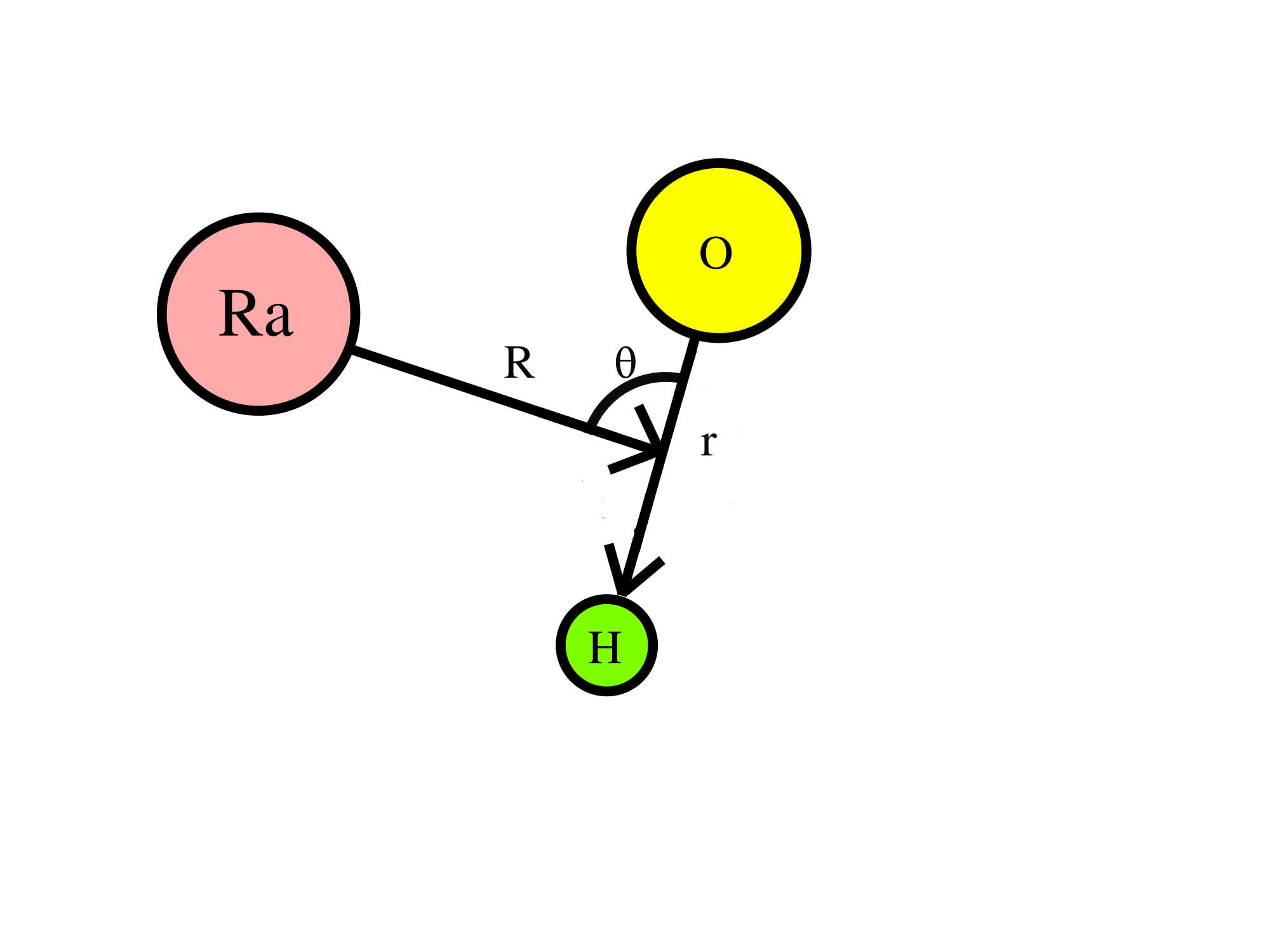}
\caption{RaOH
\label{fig:raoch3}}
\end{figure}

The theoretical description of the molecule is based on the Born-Oppenheimer approximation. The molecular wave function is the product of the electron and nuclear wave functions:
\begin{equation}
\Psi_{\rm total} \sim \Psi_{\rm nuc}(R, \hat{R}, \hat{r}) \Psi_{\rm elec}( R, \theta | q),
\label{eq:B_O_appr}
\end{equation}
where $q$ denotes the set of electron coordinates, $\hat{r}$ is the direction of the OH axis and $\hat{R}$ is the vector directed from the Ra atom to the center of mass of the ligand, see Fig.~\ref{fig:raoch3}. In the framework of this study, the ligand is considered rigid since its oscillation frequencies are significantly higher than those of the Ra—OH bond, \cite{Zakharova2026}. The geometry of the ligand is described in this way by the distance OH, $r(O-H) =1.832\;a.u.$. Both the electronic and rovibrational wave functions are thus explicitly determined by the configuration of the molecule.
%картинка нужна новая, потому что журналу может не понравится похожесть. Нужно и идейно что-то новое. Возможно, стоит сделать двойную картинку, молекула и уровни. Подумаю.

\section{Electronic wave functions}

The first step in calculating the electronic structure is to obtain molecular orbitals. The Dirac-Hartree-Fock (DHF) method is used for this purpose. Then it is necessary to take into account electronic correlations, which are addressed using the coupled cluster method (CCSD). From a technical point of view, DHF and CCSD calculations are performed within the DIRAC19\cite{DIRAC19} software package. Since the RaOH molecule has a large number of electrons due to the presence of a radium atom, the generalized relativistic effective core potential is used to reduce the effective number of orbitals. Molecular spinors (two-component orbitals) obtained using the GRECP method have incorrect behavior in the core region. Correct behavior can be restored using the one-center restoration method. Let's outline the basic idea of this method.

At the first stage, the atomic problem for Ra is solved using two different methods. First, a full-electron calculation using the Hartree-Fock-Dirac method is carried out, the application of which allows us to obtain the following set of solutions:

\begin{equation}
\Phi_{nljm}(\mathbf{r}) = \begin{pmatrix}f_{nlj}(r)\mathcal{Y}_{ljm}(\theta,\phi)\\i g_{nlj}(r)\mathcal{Y}_{2j-l,jm}(\theta,\phi)\end{pmatrix},
\label{eq:base_wf}
\end{equation}
where $\mathcal{Y}_{ljm}$ -- a set of two-component spherical spinors. Then, the GRECP method solves a molecular problem, and the resulting set of solutions has the form:

\begin{equation}
\tilde{\Phi}_{nljm}(\mathbf{r}) = \tilde{f}_{nlj}(r)\mathcal{Y}_{ljm}(\theta,\phi).
\end{equation}
The set of two-component solutions is mapped to the set of four-component solutions according to the principle of identification of quantum numbers, as a result of which equivalent bases are constructed.

At the second stage, the molecular problem is solved by the GRECP method. In the vicinity of the nucleus, the molecular orbitals $\tilde{\Psi}(\vec{r})$ are expressed as a decomposition on a two-component basis:

\begin{equation}
\tilde{\Psi}(\vec{r})=\sum_{njl} C_{njlm}\tilde{\Phi}_{njlm}(\vec{r}).
\end{equation}
The obtained coefficients are used to construct a four-component spinor correcting a two-component:
\begin{equation}
\Psi(\vec{r})=\sum_{njl} C_{njlm}\Phi_{njlm}(\vec{r}).
\end{equation}

Thus, the wave function of the electron will be decomposed according to the basis of the form ~(\ref{eq:base_wf}). By writing the wave function in the form of such a decomposition, the task of calculating two-electron matrix elements for the sensitivity parameter ~(\ref{eq:enh_param}) is greatly simplified, which will be illustrated in the corresponding section.

\section{Rovibrational wave functions}

After obtaining the wave functions of the electrons, it is necessary to calculate the rovibrational wave function $\Psi_{\rm nuc}$ from the equation (\ref{eq:B_O_appr}). It satisfies the Schrodinger equation of the form:
\begin{equation}
\hat{H}_{\rm nuc} \Psi_{\rm nuc}(R, \hat{R}, \hat{r}) = E \Psi_{\rm nuc}(R, \hat{R}, \hat{r})
\label{eq:Shreq}
\end{equation}
with an adiabatic Hamiltonian describing the motion of the nucleus. The RaOH molecule can be described as a linear rotator of length $R$ formed by the Ra atom and the center of mass of the ligand. The OH ligand is considered rigid. 
% (здесь см. чат).
% не может быть молекулой в целом, с точки зрения жесткого ротора. Просто отделили часть микрофона, и можно было бы это сделать, даже если бы наблюдались изменения формы лиганда. Отедение палки не связано с тем, что лиганд жесткий.
Such a rotator has an angular momentum of $\hat{\vec{L}}$ in the reference frame associated with the laboratory coordinate system. Let's denote the moment of the of ligand by $\hat{\vec{j}}$. Then the adiabatic nuclear Hamiltonian will be written as:

\begin{equation}
\hat{H}_{\rm nuc}=-\frac{1}{2\mu}\frac{\partial^2}{\partial R^2}+\frac{\hat{L}^2}{2\mu R^2}+\frac{\hat{j}^2}{2\mu_{OH}r^2}+V(R,\theta),
\label{eq:nuc_ham}
\end{equation}

where $\mu$ and $\mu_{OH}$ are the reduced masses of the system Ra-OH and OH, and $V(R,\theta)$ is the adiabatic electronic potential. To solve the equation~(\ref{eq:Shreq}) uses the coupled channel method developed for triatomic molecules in the work~\cite{mcguire1974quantum}. The solution is sought in the form of decomposition by eigenfunctions of the angular momentum operators:

\begin{equation}
\Psi_{\rm nuc}(R, \hat{R}, \hat{r}) = \sum_{L=0}^{L_{max}}\sum_{j=0}^{j_{max}} F_{JjL}(R)\Phi_{JjLM}(\hat{R},\hat{r}),
\label{psiexp_RaOH}
\end{equation}
where 
\begin{equation}
\Phi_{JjLM}(\hat{R},\hat{r}) = \sum_{m_L,m_j} C^{JM}_{Lm_L,jm_j} Y_{Lm_L}(\hat{R})Y_{jm_j}(\hat{r})
\end{equation}
Above, infinite summations are truncated by parameters with the index "max"{}. Thus, we have obtained a system of equations for the radial part of $F_{JjL}(R)$:

\begin{equation}
\Big[\frac{d^2}{dR^2}-\frac{L(L+1)}{R^2}-\frac{\mu j(j+1)}{\mu_{OH} r^2}+2\mu E\Big]F_{JjL}(R)
=2\mu\sum_{\tilde{j},\tilde{l}}\mathcal{V}_{jL,\tilde{j}\tilde{L}}^J F_{J\tilde{j}\tilde{L}}(R),
\end{equation}
where the matrix potential is defined in the right part
\begin{align}
\mathcal{V}_{jL,\tilde{j}\tilde{L}}^J(R)=(-1)^{J+j+\tilde{j}}\sqrt{(2L+1)(2\tilde{L}+1)(2j+1)(2\tilde{j}+1)}\nonumber\\
\times \sum_{k=0}^{k_{max}}V_{k}(R) \begin{Bmatrix}j&L&J\\\tilde{L}&\tilde{j}&\lambda\end{Bmatrix}
\begin{pmatrix}L&\lambda&\tilde{L}\\0&0&0\end{pmatrix}\begin{pmatrix}j&\lambda&\tilde{j}\\0&0&0\end{pmatrix}
\end{align}
where $V_k(R)$ they correspond to the decomposition of the potential by Legendre polynomials $P_k(x)$:
\begin{equation}
V(R,\theta) = \sum_{k=0}^{k_{max}}V_k(R) P_k(\cos\theta).
\end{equation}
This equation is solved using the finite basis method with a basic set composed of the harmonic oscillator's eigenfunctions using a locally developed C++ software package.

\section{Calculation of the sensitivity parameter}
\label{sec:PWE}

Partial decomposition is used to calculate matrix elements of the form (\ref{eq:ax_two_el_int}):
\begin{equation}
\frac{1}{r_{12}} = \sum_{L=0}^{\infty} \sum_{M=-L}^{L} \left( \frac{4 \pi}{2L+1} \right) \left( \frac{r^L_<}{r^{L+1}_>} \right) Y_{LM}^* (\hat{r}_1) Y_{LM} (\hat{r}_2),
\label{PE}
\end{equation}
where $r_> = \max (\vec{r}_1, \vec{r}_2 )$ и $r_< = \min (\vec{r}_1, \vec{r}_2)$. 
The explicit form of the angular dependence of the electronic wave functions is known, see equation ~(\ref{eq:base_wf}). By performing a number of laborious algebraic transformations and analytically integrating over the angles, the expression ~(\ref{eq:ax_two_el_int}) can be reduced to the form:

\begin{equation}
\begin{gathered}
A(a, b | c, d) = \sum_{JM} (-1)^ {j_{a}-m_{a} + J - M + j_{b} - m_{b}}
\times \\ \times
{\begin{pmatrix}j_a&J&j_c\\-m_{a}&M&m_{c} \end{pmatrix}}
{\begin{pmatrix}j_b&J&j_d\\-m_{b}&-M&m_{d} \end{pmatrix}} \langle a b || \frac{1}{r_{12}} || c d\rangle_J^A.
\end{gathered}
\label{M_el_expansion}
\end{equation}
The matrix element in the right part is equal to
\begin{equation}
\begin{gathered}
\langle ab || \frac{1}{r_{12}} || cd \rangle_J^A = - \int_0^\infty dr_{1} dr_{2} (-1)^{J} G_{J}(\kappa_{a},-\kappa_{c}) G_J ( \kappa_{b}, \kappa_{d} )
\times \\ \times
g_J(r_1, r_2) B_{ac} (r_1) C_{bd} (r_2).
\label{M_el_a}
\end{gathered}
\end{equation}
The function $g_J$ in the above expressions arises naturally from the partial decomposition~(\ref{PE}):
\begin{equation}
\begin{gathered}
g_{J} (r_1, r_2) = \frac{r^L_<}{r^{L+1}_>}.
\end{gathered}
\end{equation}
The coefficients, which include the radial part of the wave functions, have the form:
\begin{equation}
\begin{gathered}
A_{ab} (r_1) = G_{a} (r_{1}) G_{b}(r_{1}) + F_{a}(r_{1}) F_{b} (r_{1}), \\
B_{ab} (r_1) = G_{a} (r_{1}) F_{b}(r_{1}) + F_{a}(r_{1}) G_{b} (r_{1}), \\
C_{ab} (r_1) = G_{a} (r_{1}) G_{b}(r_{1}) - F_{a}(r_{1}) F_{b} (r_{1}), \\
G_{a}(r) = r g_a(r), \quad F_{a}(r) = r f_a(r).
\end{gathered}
\end{equation}
The angular coefficient of $G_J$ is:
\begin{equation}
\begin{gathered}
G_{J}(\kappa_{a},\kappa_{b}) = (-1)^{j_b + 1 \slash 2} \sqrt{[j_a][j_b][l_a][l_b]} {\begin{pmatrix}l_a&J&l_b\\0&0&0\end{pmatrix}} {\begin{Bmatrix}j_a&J&j_b\\l_b&\frac{1}{2}&l_a\end{Bmatrix}},
\end{gathered}
\end{equation}
where $[a] = 2 a + 1$. 

A Python software package has been developed to calculate two-electron integrals. Radial integrals are calculated using Simpson quadratures. It should be noted that the function $g_J$ has a discontinuity of the derivative with matching arguments, and numerical integration over $r_1$ and $r_2$ should be performed with caution.

In the SCF approximation the electronic wavefunction can be represented as,
\begin{equation*}
\Psi_{\rm el}(x_1,\ldots x_N)\simeq \frac{1}{\sqrt{N!}}\begin{vmatrix}
\psi_1(x_1)\ldots \psi_N(x_1)\\
\ldots
\psi_1(x_N)\ldots \psi_N(x_N
\end{vmatrix}
\end{equation*}
Where $x_k$ denotes both spatial and spin coordinate. 
If we further restrict ourselves to a Kramers-restricted SCF (as in Dirac program suite) the occupied orbitals are divided into pairs of orbitals related by $\mathcal{T}$ transform forming closed shells, and a single unpaired open-shell orbital, which we will denote as $n$.
Then taking into account that the interaction is $\mathcal{T}$-odd with respect to the state of the first electron, and $\mathcal{T}$-even with respect to the state of the second electron, the interaction Hamiltonian averaged over the electronic wavefunction can be obtained as,
\begin{equation}
\langle \Psi_{el}| \hat{H}_{ee} |\Psi_{el}\rangle \simeq \sum_{m\in \text{closed shells}} \Big( A(n m| m n) - A(n m | m n) - A( m n | n m)\Big)
\end{equation}

The results are summarized in the Table \ref{tbl:table1}

\begin{table}[h]
\small
  \caption{Electron-electron axion-mediated interaction}
  \label{tbl:table1}
  \renewcommand{\arraystretch}{1.5}
  \begin{tabular*}{0.7\textwidth}{@{\extracolsep{\fill}}llll}
    \hline\hline
$v_{\parallel}$ & $v_\perp$ &  $l$ &  $W_{ee}$,  $10^{-5}\lambda_e^{-1}$\\
\hline
  \multicolumn{4}{l}{Equilibrium configuration}\\
\hline
 &  &  & 1.383    \\
\hline
  \multicolumn{4}{l}{J=0}\\
\hline
0 & 0 & 0 & 1.376 \\
1 & 0 & 0 & 1.378 \\
2 & 0 & 0 & 1.379 \\
0 & 2 & 0 & 1.356 \\
\hline
  \multicolumn{4}{l}{J=1}\\
  \hline
0 & 0 & 0 & 1.376 \\

0 & 1 & 1 & 1.367 \\

\hline\hline
  \multicolumn{4}{l}{The YbOH equilibrium (FS-CCSD, $m_a=1..10$eV) \cite{maison2021electronic}}\\
  \hline
&  &  & 1.46    \\
 \hline\hline
  \multicolumn{4}{l}{The BaF equilibrium  (CCSD, $m_a=1..10^2$eV) \cite{prosnyak2024axion}}\\
  \hline
&  &  & 0.8  \\
 %\hhline{===}
 \hline\hline
  \end{tabular*}
\end{table}

\section{Acknowledgments}
The work was supported by the Russian Science Foundation (grant number 24-72-00002, https://rscf.ru/project/24-72-00002/).

\bibliographystyle{unsrt}
%\bibliography{RaOCH3, Zakharova}

\end{document}